\documentclass[aps,pra,reprint]{revtex4-1}
\usepackage{graphicx}
\usepackage{amsmath}
\usepackage{amsfonts}
\usepackage[utf8]{inputenc}
\usepackage{xcolor}
\usepackage{physics}
\usepackage{overpic,color}
\usepackage{hyperref}

\newcommand{\twistwp}{\psi_{s}}
\newcommand{\posv}{\vec{r}}

\newcommand{\rt}{(\posv,t)}
\newcommand{\volact}{S_{\omega_0}}
\newcommand{\volquiv}{z_{\omega_0}}

\newcommand{\volkovwp}{\psi_\vec{k}}
\newcommand{\ionpot}{I_{\text{p}}}
\newcommand{\pondpot}{U_{\text{p}}}

\newcommand{\rhomax}{\rho_{\text{max}}}
\newcommand{\wmax}{\omega_{\text{max}}}
\newcommand{\wmin}{\omega_{\text{min}}}

\newcommand{\aeff}{a_{\text{eff}}}
\renewcommand{\vec}{\mathbf}
\newcommand{\ii}{\mathrm i}

\definecolor{custompurple}{rgb}{1,0,1}
\definecolor{customgreen}{HTML}{aadc32}
\definecolor{customdarkgreen}{HTML}{6d9311}

\begin{document}
\title{High Harmonic Generation with Twisted Electrons}
\author{Sebastian Gemsheim}
\email{sebgem@pks.mpg.de}
\author{Jan-Michael Rost}
\affiliation{Max Planck Institute for the Physics of Complex Systems, N\"othnitzer Stra\ss e 38, D-01187 Dresden, Germany}

\begin{abstract}
We present analytically and numerically the spectrum of high harmonic emission generated by twisted electrons in the presence of linearly polarized light. %
Ensuing transitions from electronic continuum states with orbital angular momentum to bound states give rise to circularly polarized attosecond pulses. For central collisions with twisted wavepackets  continuum-bound transitions are subject to dipole selection rules. For non-central collisions a crossover from circularly to linearly polarized emission occurs for increasing impact parameter, due to the transverse topology of twisted wavepackets.
\end{abstract}

\maketitle
Beams can have a  transverse spatial profile which allows them to carry orbital angular momentum (OAM). It appears as an azimuthal phase dependence in the plane perpendicular to the beam direction resulting in a helical phase front giving rise to the name twisted beams. Following theoretial proposals, apart from light~\cite{bava+90}, the most obvious realization of beams,  also electrons~\cite{ucto10, veti+10} and (possibly~\cite{caja+18}) neutrons \cite{clba+15} have been  experimentally realized as twisted beams due to the wave nature of quantum mechanics.

\section{Introduction}
Recent efforts aim at a controlled production of short and  intense XUV  light pulses through high harmonic generation (HHG)~\cite{hepi+13,pabo+18} driven by a twisted light beam. The underlying mechanism is the nonlinear interaction of the twisted (or otherwise sculpted) light beam with an electron in the presence of an ion leading to the up-conversion of incident photons with  frequency $\omega_{0}$ to much higher frequency  $n\omega_{0} \gg \omega_{0}$  by recombination of electrons with the ion, typically described in three steps~\cite{Kulander1993}. In the first step, the electron wave packet is generated through ionization with the incident twisted light beam and possesses a specific initial phase depending on its azimuthal position within the laser field. After propagation in the continuum (step 2) a small part of the electron wavepacket recombines (step 3) under the emission of a single photon. The macroscopic superposition of the emitted, linearly polarized HH radiation from the target, e.g., an ensemble of atoms, gives then rise to a light vortex with an up-converted OAM proportional to the harmonic order of the emitted frequency. In the processes described the emphasis is on the  question which input of light produces which kind of harmonic generation macroscopically.

\begin{figure}
	\begin{center}
		\begin{overpic}[width=0.5\textwidth]{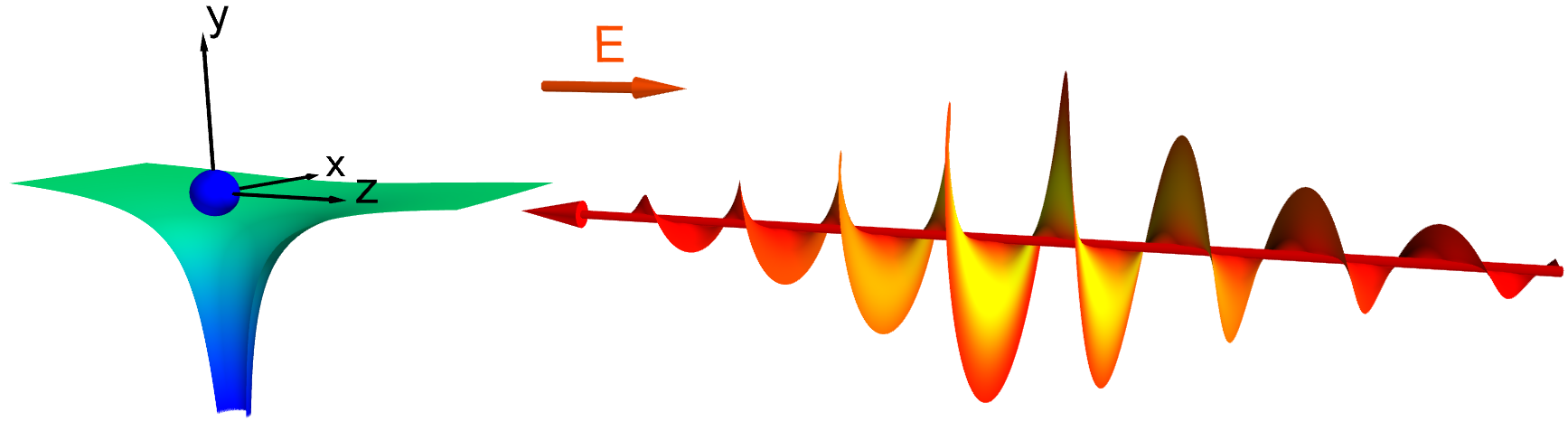}
			\put(15,19){\color{blue}\Large $\textbf{H}^+$}
			\put(62,23){\color{red}\Large $\textbf{e}^-$}
		\end{overpic}
	\end{center}
	\caption{Laser assisted scattering of a singly charged positive ion with a  twisted electron  in a laser field. The electric field is polarized in the $z$-drection.}
	\label{fig:collision_scenario}
\end{figure}

In the following we aim at a microscopic understanding of the basic nonlinear HHG process involving OAM on the level of a single emitter. In particular we ask, which kind of high harmonics is produced, if the vortex character is imprinted on the electronic wavepacket~\cite{Bliokh2007,Bliokh2017} instead of the light beam. To this end we will collide a well defined twisted electron wavepacket with an ion under (standard) linearly polarized light, see Fig.~\ref{fig:collision_scenario}. This process can be viewed  as a laser-assisted collision~\cite{saro99}. Guided by work with standard Gaussian electrons~\cite{Zagoya2012,Zagoya2012a}, we will be able to provide an analytical approximation allowing us to understand this non-linear process and its basic features with their parameter dependence. 

\section{Collision of a twisted electron beam with an ion under laser illumination}

\begin{figure}
	\begin{center}
	\includegraphics[width=0.5\textwidth]{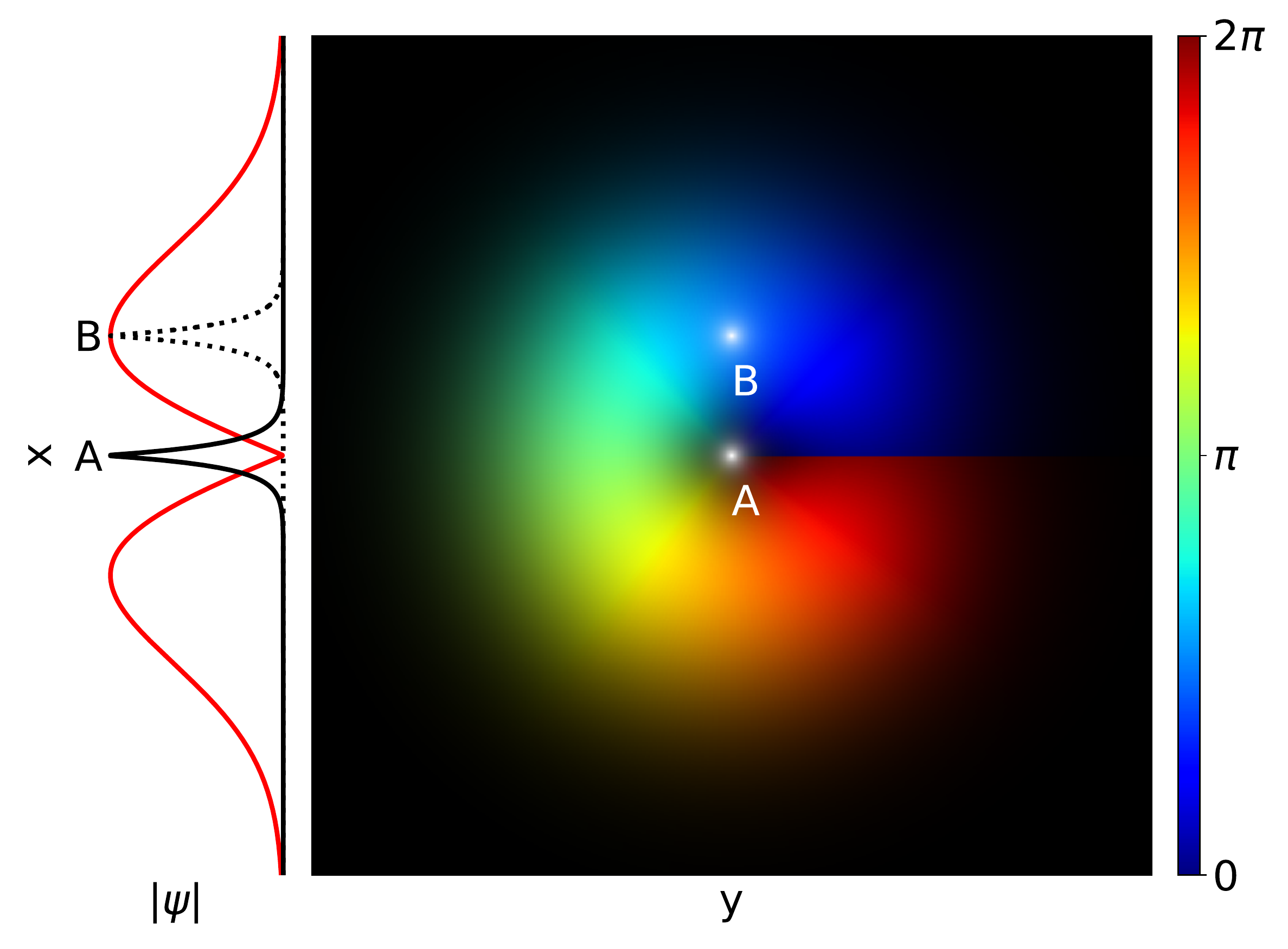}
	\end{center}
	\caption{Transverse topology of a twisted Volkov wavepacket (colored) with $s=1$ and the hydrogen ground state (white). A central~(A, $b=0$ in \eqref{eq:hamiltonian}) and non-central~(B, $b\ne0$) impact is shown. The left panel shows a cross section through the transverse origin and the different overlaps for both scenarios. In the right panel the wavefunction moduli $\abs{\psi_{s=1}(x,y,z=z_{\omega_0}(t),t)}$ and $\abs{\psi_{100,b}(x,y,z=0,t)}$, respectively, are represented by the transparency. Additionally, the phase of $\twistwp$ is color-coded from $0$ to $2\pi$.}
	\label{fig:twisted_topo}
\end{figure}

We consider a single active electron in a non-relativistic framework  exposed to an ion at position $(b,0,0)$ with Coulomb potential $V_{b}(\vec{r})=-1/|\vec{r}-b\vu{x}|$  and a linearly polarized light field along the $z$-axis described in  dipole approximation by the classical vector potential $\vec{A}(t)=-(E_0/\omega_0) \vu{z} \sin(\omega_0 t)$. It also determines the average energy of the electron in the laser field (ponderomotive potential)
 $U_\mathrm{p}= A_{0}^{2}/4 \equiv  (E_0/2\omega_0)^{2}$ and the maximal excursion of the electron in the laser field (quiver amplitude) $d_{\omega_{0}}\equiv E_{0}/\omega_{0}^{2}$. The time-dependent Hamiltonian for the electron reads  (atomic units are used throughout unless noted otherwise)
\begin{equation}
	{H_{b,A}}(t) = -\frac{1}{2} \nabla^{2}_{\vec r} + V_{{b}}(\vec{r}) + V_{{A}}(t,z) \,,	\label{eq:hamiltonian}
\end{equation}
where the light-matter coupling $V_{A}(t,z) = - {z}\cdot \partial_t A(t)$ is expressed in length gauge.
We take the twisted electron wavepacket  initially at $t=0$
localized at the quiver distance  $ d_{\omega_{0}}\gg 1$ away from the nucleus where
$V_{A}\gg V_{b}$. Hence, it can be constructed from a superposition of Volkov states, 
$\volkovwp\rt \propto \exp\left(\ii(\vec k +\vec A(t))\cdot\vec{r} - \frac\ii 2\int_0^t \dd{t'} (\vec k +\vec A(t'))^2\right)$, which are solutions of the Schr\"odinger equation with $V_{b}=0$ in \eqref{eq:hamiltonian},  describing  free electrons in a laser field~\cite{Joachain2012}. The momentum weights $w_{s}(\vec k)$  with vortex charge $s$ carry the twisted electron property  which we encode in a Laguerre-Gaussian (LG) shape~\cite{Bialynicki-Birula2017} and width $\beta$, expressed
in cylindrical coordinates $(k_{\rho}, \phi, k_{z})$ with the axis of OAM quantization and  laser polarization chosen to be collinear:
\begin{equation}
	w_{s}(\vec k) \propto e^{\ii s\phi} k^s_{\rho} \exp \left[ -\beta^2 \left( k^2_{\rho} + k^2_{z} \right) / 2 \right] \,.
\end{equation}
 Due to its invariance under a Hankel transform, the resulting spatial twisted Volkov wavepacket  is also of LG form in position representation $(\rho, \varphi, z)$,
\begin{subequations}
\begin{align}
	\twistwp \rt &= \int\dd[3]{\vec k} w_{s}(\vec k) \cdot \volkovwp(\vec{r},t) \nonumber \\
	&= N_{s}(t) \exp\left[\ii( s\varphi+A(t) z -\volact(t)) \right]  \nonumber \\
	&\quad \times \rho^s \exp\left[ - \frac{\rho^2 + (z - \volquiv(t))^2}{2\sigma^2(t)}   \right]    				\label{eq:twisted_wp}
\end{align}
with normalization constant
\begin{equation}
	N_{s}(t) \equiv \left( \beta / \abs{\sigma^2(t)} \right)^{(s+3/2)} / \left( \pi^{3/4} \sqrt{s!} \right) \,.
\end{equation}
\end{subequations}
For $s=0$ one retrieves the familiar Gaussian wavepacket with width $\sigma(t) \equiv  \sqrt{\beta^2+it}$ whose center follows a classical trajectory in the laser field
$\vec r_{\omega_{0}}(t)=(0,0,z_{\omega_{0}}(t))$ with $z_{\omega_{0}(t)} = 
d_{\omega_{0}}\cos(\omega_0t)$ and  action $\volact(t)\equiv \pondpot \left[ t - \sin(2\omega_0 t)/(2\omega_0) \right]$. Obviously a twisted wavepacket with $s\ne 0$ retains 
this time evolution but acquires the typical transverse ring structure with an amplitude maximum at $\rhomax^{(s)}(t) \equiv \sqrt{s}\beta\sqrt{1+t^2/\beta^4}$ which renders it  an  OAM eigenstate ($\hat{L}_z\ket{\twistwp} = s \ket{\twistwp}$). The transverse topology is illustrated in Fig.~\ref{fig:twisted_topo}.

In addition, the first exponential term in Eq.~(\ref{eq:twisted_wp}) describes the helical phase front common to all propagating twisted wavepackets. Due to the chosen geometry, the vortex charge is not altered by the electric field. Even the inclusion of the magnetic field component via a first order $1/c$-expansion of the full vector potential $\vec{A}\left( w_0 \cdot \left[ t-\vec{e}_x\cdot\vec{r}/c \right] \right)$~\cite{Brennecke2018} only results in a drift $\expval{\hat{y}}(t) = \volact(t)/c$ perpendicular to the quantization axis and also conserves the vortex charge $s$. (Note that  the vacuum speed of light is  $c\approx 137$ in atomic units). HHG without OAM is recovered for an unstructured Gaussian wavepacket.

\section{Harmonic emission}

Since we are interested in the highest harmonic frequencies in the emission spectrum, we consider only recombination to  the ground state. Other recombination channels are at least an order of magnitude weaker in harmonic yield, see Appendix~\ref{app:hhg_yield}. In the low-frequency approximation and in the quasi-static limit at the instantaneous electric laser field the transition element $\vec{M}^{(s)}_{b}$ of a laser-assisted radiative recombination can be approximated~\cite{Bivona2005} by 
\begin{equation}
	\vec M^{(s)}_{b}(t) \equiv \mel{\psi_{100,b}(t)}{\vec p}{\psi_s(t)} \label{eq:transition_el}\,,
\end{equation}
where the momentum operator is denoted by $\vec p$. The interaction potential is weak at the origin ($V_{b}\gg V_{A}$) and the highest harmonics are emitted when the electric field is close to zero. Therefore, we neglect the influence of the laser on the ground state and use the unperturbed hydrogen ground state $\psi_{100,b}\rt = \exp\left(-\ii E_{1} t\right)\exp\left(-\abs{\vec{r}-b\vu{x}}\right)/\sqrt{\pi}$ with $E_{1}=-1/2\equiv -I_p$. Likewise, we neglect the influence of the Coulomb potential on the twisted Volkov wavepacket. Even though the Coulomb potential $V_{b}$ is stronger than the laser coupling term in the vicinity of the ion, the accelerated twisted wavepacket spends only a short amount of time $t\simeq \mathcal{O}(1/A_0)$ close to the ion rendering the influence of $V_{b}$ small.  The corresponding emission spectrum reads

\begin{subequations}
\label{eq:spectrum}
\begin{equation}
	\mathcal{S}^{(s)}_{b}(w,T) = \sum_{j={x,y,z}} \mathcal{S}^{(s)}_{j,b}(w,T) 
\end{equation}
for an interaction time $T$, where 
\begin{equation}
	\mathcal{S}^{(s)}_{j,b}(w,T) = \frac{\omega^2}{2\pi} \abs{\int_0^T \dd{t} e^{\ii\omega t} \, M^{(s)}_{j,b}(t) }^2 
\end{equation}
\end{subequations}
with vector components $M^{(s)}_{j,b}$. In this work we apply an additional time window in the Fourier transform for the reduction of edge effects, see Appendix~\ref{app:ft_filter}

\subsection{Central collisions}

For central collisions $b=0$ the transition element~(\ref{eq:transition_el}) introduces dipole selection rules, familiar from atomic bound-bound transitions. Hence, recombinations with $\lvert s \rvert > 1$ are suppressed.  The $z$-component of Eq.~(\ref{eq:transition_el}) is only non-vanishing for a Gaussian Volkov wavepacket ($s=0$) and the transverse components are only non-zero for $s=\pm1$. Therefore, we set $s=1$ to allow recombination to the ground state. Under the emission of a single photon higher vortex charge states can only recombine into excited bound states with magnetic quantum number $m = s,s\pm 1$. The analytical approximation (for $\beta\gg 1$) for the transition element reads
\begin{equation}
	\vec{M}^{(1)}_{b=0}(t) =  \begin{pmatrix}
	1 \\ \ii \\ 0
	\end{pmatrix}
	\mathcal{N}(t) \frac{ e^{-\ii\left( \volact(t) + \ionpot t  \right)}   }{\left[ 1 + a(t)^2 \right]^2 }   \exp \left[-\frac{\volquiv^2(t)}{2\sigma^2(t)} \right] \,	\label{eq:ground_trans_elem}
\end{equation}
with $a(t) \equiv A(t) - \ii \volquiv(t) / \sigma^2(t)$ and $\mathcal{N}(t) \equiv 8 \sqrt{\pi} N_{s}(t)$.
The dipole response in the direction of the electric field vanishes in this process, contrary to HHG with $s=0$, where~(\ref{eq:transition_el}) has only a contribution parallel to the polarization axis $\vu{z}$. The two transverse components differ by a phase  of $\exp(\ii\pi/4)$. The physical interpretation of this emission pattern is that of a dipole rotating in the transverse plane instead of a linearly oscillating dipole in the longitudinal direction of the electron motion for $s=0$. This corresponds to the emission of high-energetic, circularly polarized pulses, due to the transfer of angular momentum, which is not present in HHG with $s=0$. Moreover, the direction of maximum emission will point parallel to the polarization axis $\vu{z}$, in contrast to linearly polarized HHG. This is reminiscent of the emission from atomic bound-bound transitions with magnetic quantum number difference $\abs{\Delta m}=1$~\cite{Fowles1989}. 

\begin{figure}
	\begin{center}
	\begin{overpic}[width=0.5\textwidth,grid=False]{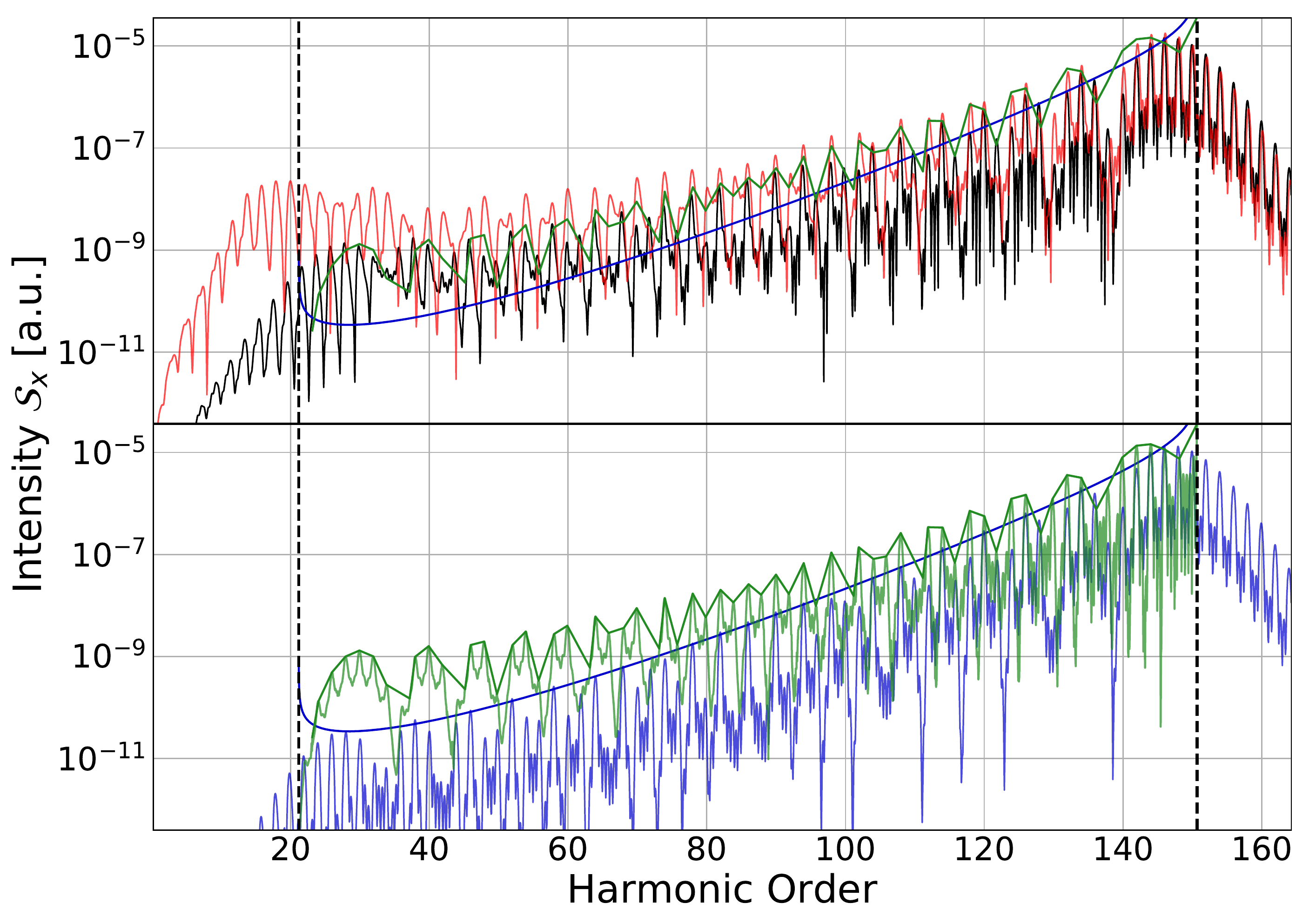}
			\put(12,65){\textbf{(a)}}
			\put(12,34){\textbf{(b)}}
		\end{overpic}
	\end{center}
	\caption{Spectrum $\mathcal{S}^{(s=1)}_{x,0}$ of the HHG emission from the recombination of a twisted electron wavepacket with the hydrogen ground state for laser field strength of $E_0 = 0.05835$, frequency $\omega_0 = 0.02360$,
	interaction time of two optical cycles, $T = 2T_{0}$, and wavepacket width $\beta = 25$. (a) The analytical result is represented by the solid black line. The red curve is the corresponding result from the numerical integration of the Schr\"odinger equation with the full Hamiltonian~(\ref{eq:hamiltonian}). (b) The SPA of the analytical result including peak envelope is shown in green and the analytical result with constant width and corresponding SPA scaling Eq.~(\ref{eq:spa_scaling}) is blue. The scaling is multiplied by a numerical constant to match the peak values. Both envelopes are also shown in (a) for comparison. Black dashed lines show the HHG plateau cutoffs $I_p/\omega_0$ and $(I_p+2U_p)/\omega_0$, respectively.}
	\label{fig:spectrum}
\end{figure}

Yet, as in the $s=0$ case, the pulse duration of the highest harmonic is on the attosecond scale for infra-red driving fields. Also the high (low) frequency cutoff $\wmax=\ionpot + 2\pondpot$ ($\wmin=\ionpot$) is unchanged by the electron OAM as Fig.~\ref{fig:spectrum} reveals, where
the HHG spectrum from recombination into the ground state using \eqref{eq:ground_trans_elem} in \eqref{eq:spectrum} is compared to the numerical solution of the Schr\"odinger equation with the full Hamiltonian Eq.~(\ref{eq:hamiltonian}) and $\ket{\psi_{s=1}(0)}$ as initial condition (see Appendix~\ref{app:num_sim}). Evaluating the Fourier integral of Eq.~(\ref{eq:spectrum}) in stationary phase approximation (SPA) we find the explicit expression
\begin{equation}
	\mathcal{S}(\omega) \propto \frac{\omega^2 \exp \left[ 2\left( \omega - \wmax \right) /(\omega_0\beta)^2 \right] }{2\omega_0 \sqrt{(\wmax-\omega) (\omega-\wmin)} \left( 1 + 2 (\omega-\wmin) \right)^4}	\,,	\label{eq:spa_scaling}
\end{equation}
which describes the envelope shape of the spectrum for constant width $\sigma(t)\approx\beta$ quite well (see Fig.~\ref{fig:spectrum}b).

The interaction time $T=2T_0$ of two laser cycles is long enough to resolve the high harmonic peaks and this particular choice of $\beta$  leads to a notable slope of the plateau between the cutoffs. Within the constant width approximation, a Jacobi-Anger expansion reveals the spectral peaks to appear at frequencies $\omega_{\mathrm{peaks}} = \ionpot + \pondpot + 2\omega_0 j$~\cite{vandeSand2000} for $j \in \mathbb{Z}$ and $\abs{j} \leq \pondpot/(2\omega_0)$.

We attribute the larger yield from the numerical solution at low frequencies in Fig.~\ref{fig:spectrum} to Coulomb focusing of the twisted wavepacket (see Appendix~\ref{app:num_sim}).

\subsection{Off-center collisions}

Next, we briefly discuss off-axis scattering with $b>0$, where the rotational symmetry of the system around the $z$-axis is broken~(Fig.~\ref{fig:twisted_topo}, B). For large $b$ the phase structure of the free electron wavepacket will not be probed anymore by the bound state and we expect a transition to HHG mainly without transfer of angular momentum.

\begin{figure}
	\begin{center}
	\includegraphics[width=0.5\textwidth]{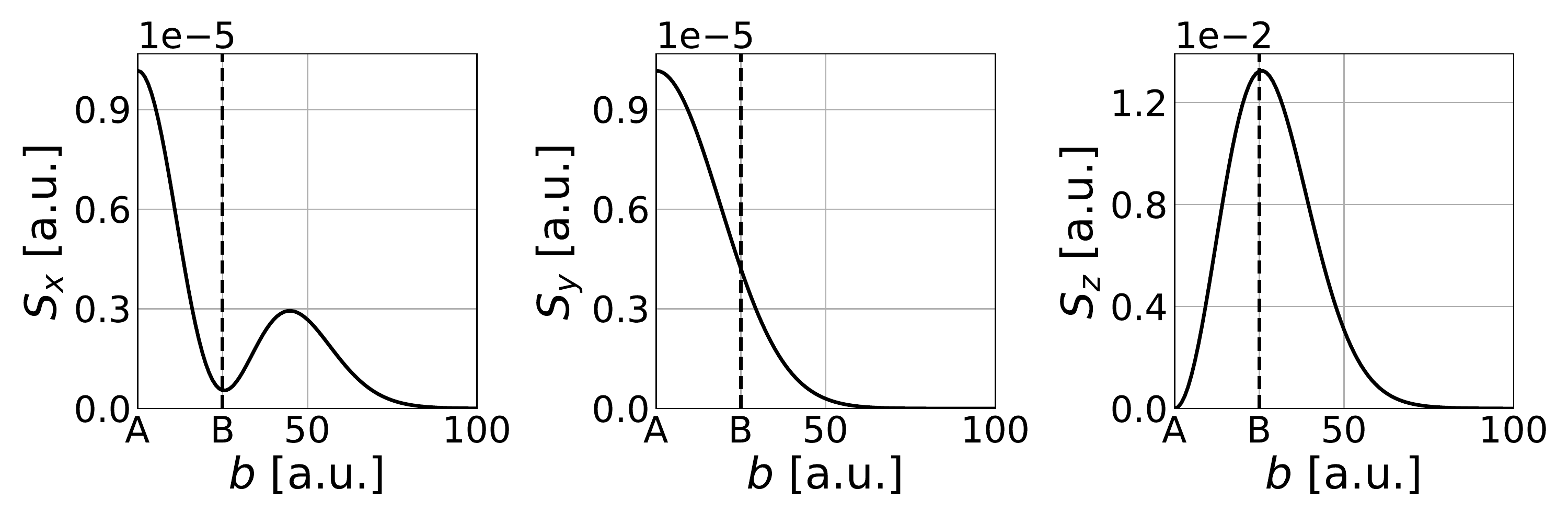}
	\end{center}
	\caption{Spectral intensities $\mathcal{S}^{(s=1)}_{i,b}$ ($i=\{x,y,z\}$) of the highest harmonic frequency in the plateau region for ground state recombination with varying impact parameter $b$. The initial transverse maxima $\rhomax^{(1)}(0)=\beta$ is indicated by the vertical, dashed line and the two scenarios A and B from Fig.~\ref{fig:twisted_topo} are marked on the horizontal axis.}
	\label{fig:spectrum_impact}
\end{figure}

The numerical results for the harmonic frequency $\omega \in [\wmax-2\omega_0,\wmax ]$ with the largest intensity are shown in Fig.~\ref{fig:spectrum_impact} for ground state recombination. 
Due to the broken symmetry, the $z$-component of the spectral intensity $\mathcal{S}^{(1)}_{z,b}$ becomes non-zero and exhibits a peak at $b \approx \rhomax^{(s=1)}(0)$ for $\omega\approx \wmax$. The reasons are an increased wavefunction overlap in the transverse plane and unequally weighted contributions from different azimuthal phases (Fig.~\ref{fig:twisted_topo}, B), resulting in elliptically polarized emission. Its amplitude is a  few orders of magnitude larger than for the other two components. The transverse components show the expected decline for large $b$. However, while the $y$-component shows a monotonically decreasing behavior, the $x$-component has a minimum at $b\approx\rhomax^{(s=1)}(0)$ and a subsequent maximum before steadily declining. Hence, three distinct regions can be identified. The emission characteristics change from a rotating dipole ($b\approx0$) over a dominant linear dipole for $b\approx\rhomax^{(s=1)}(0)$ to emission without resolved internal structure for larger impact parameters $b\gg\rhomax^{(s=1)}(0)$ where only the exponentially decaying parts of the wavefunctions overlap.

In general $b\ne 0$ removes the symmetry imposed dipole selection rules and therefore allows for a greater number of recombination channels.  A numerical analysis of the HHG yield for individual channels including excited bound states is provided in Appendix~\ref{app:hhg_yield}.

\section{Direct probe of angular momentum transfer}
\subsection{A proposal for an experiment}

\begin{figure}
	\begin{center}
	\begin{overpic}[width=0.4\textwidth,grid=False]{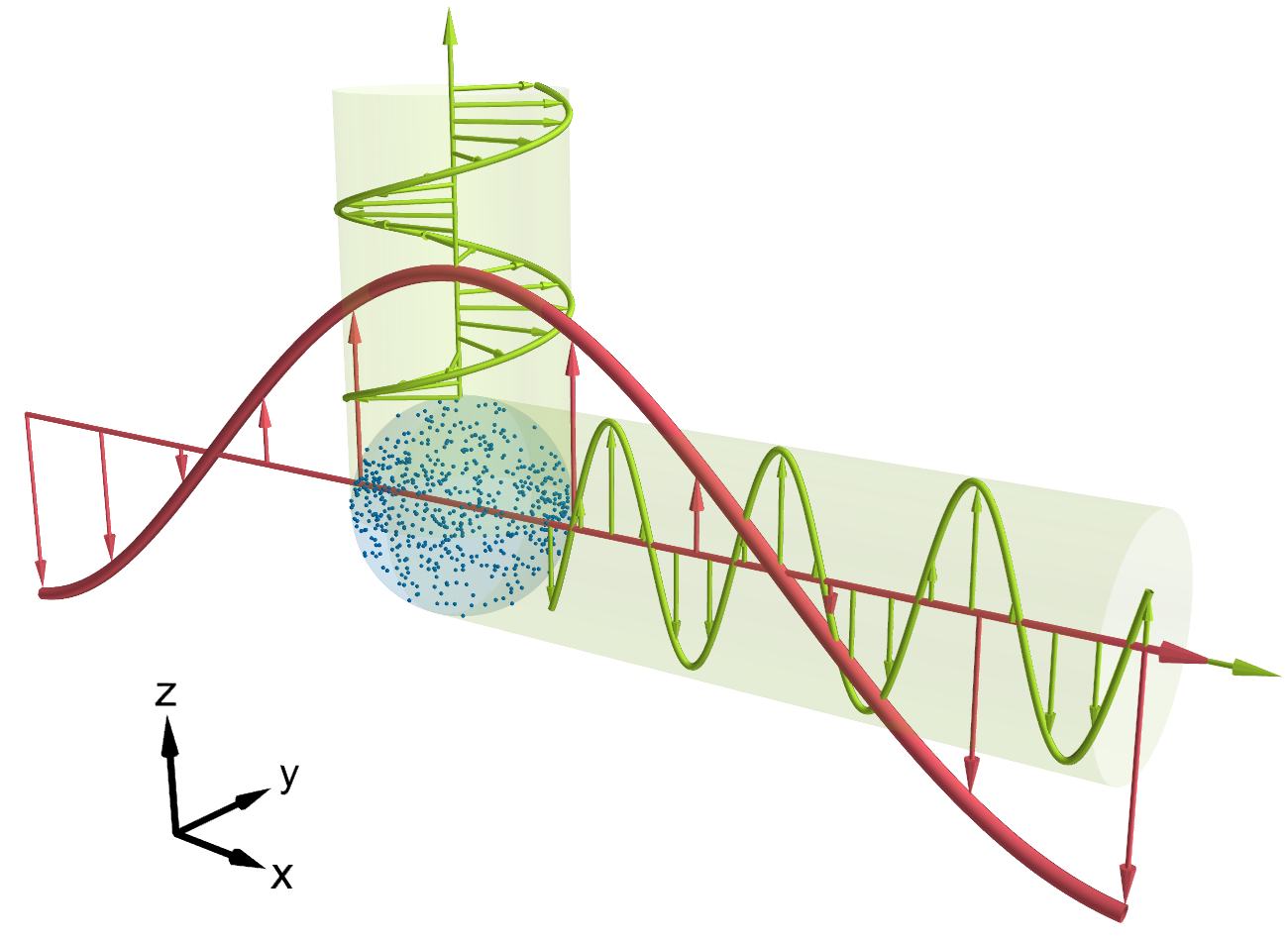}
			\put(70,38){\color{customdarkgreen}\Large $\mathbf{s=0}$}
			\put(7,55){\color{customdarkgreen}\Large $\mathbf{s=1}$}
		\end{overpic}
	\end{center}
	\caption{Comparison of emission directions for macroscopic HHG from an ensemble of atoms (blue) for $s=0,1$. For $s=0$, the HH emission (green) propagates in the same direction as the driving field (red) and has the same linear polarization. In contrast, the circularly polarized HH radiation from $s=1$ is emitted in the polarization direction, perpendicular to the laser propagation axis.}
	\label{fig:emission_direction}
\end{figure}

From the results it has become clear that it is the azimuthal dependence $s$ which a twisted electron wavepacket
transfers to the high harmonics in form of circularly polarized light. Hence, we conclude with the proposal for an experiment probing this transfer of angular momentum directly: First, atoms with  an $s$-state valence electron are prepared in a $p$-state with $m=1$ via coherent pumping with a resonant circularly polarized laser pulse with propagation in $z$-direction. Subsequently, the system is illuminated by a strong $\vu{z}$-linearly polarized infra-red laser pulse that propagates in the $x$-direction. The second laser pulse cannot induce angular momentum transfer and the ionized wavepacket will possess the vortex factor $e^{\ii\varphi}$. The re-collision will then be analogous to the situation analyzed here and the impact parameter will be close to zero if the magnetic field strength is negligible. A similar setting was used in Ref.~\cite{Xie2008}, where, however, the laser polarization is orthogonal to the OAM (in the $z$-direction) of the excited state. The helicity of the emitted high harmonics depends solely on the helicity of the weak pump pulse. A positive side effect is the possibility to effectively adjust phase-matching, even at high levels of ionization, along the HH propagation direction in a gas medium, necessary for the coherent build-up of high harmonic radiation~\cite{Popmintchev2012,Hernandez-Garcia2010}. It can be achieved with a macroscopically, $z$-dependent phase delay $t_{\text{delay}}(z) = t - z \cdot n_{\text{delay}} / c$ of the second laser pulse front with freely tunable parameter $n_{\text{delay}}$, due to the orthogonal emission direction (Fig.~\ref{fig:emission_direction}). Also, the standard HHG from the remaining ground state atoms does not harm since it is emitted perpendicularly.

\subsection{Illustrative  example}
\begin{figure}
	\begin{center}
	\includegraphics[width=0.4\textwidth]{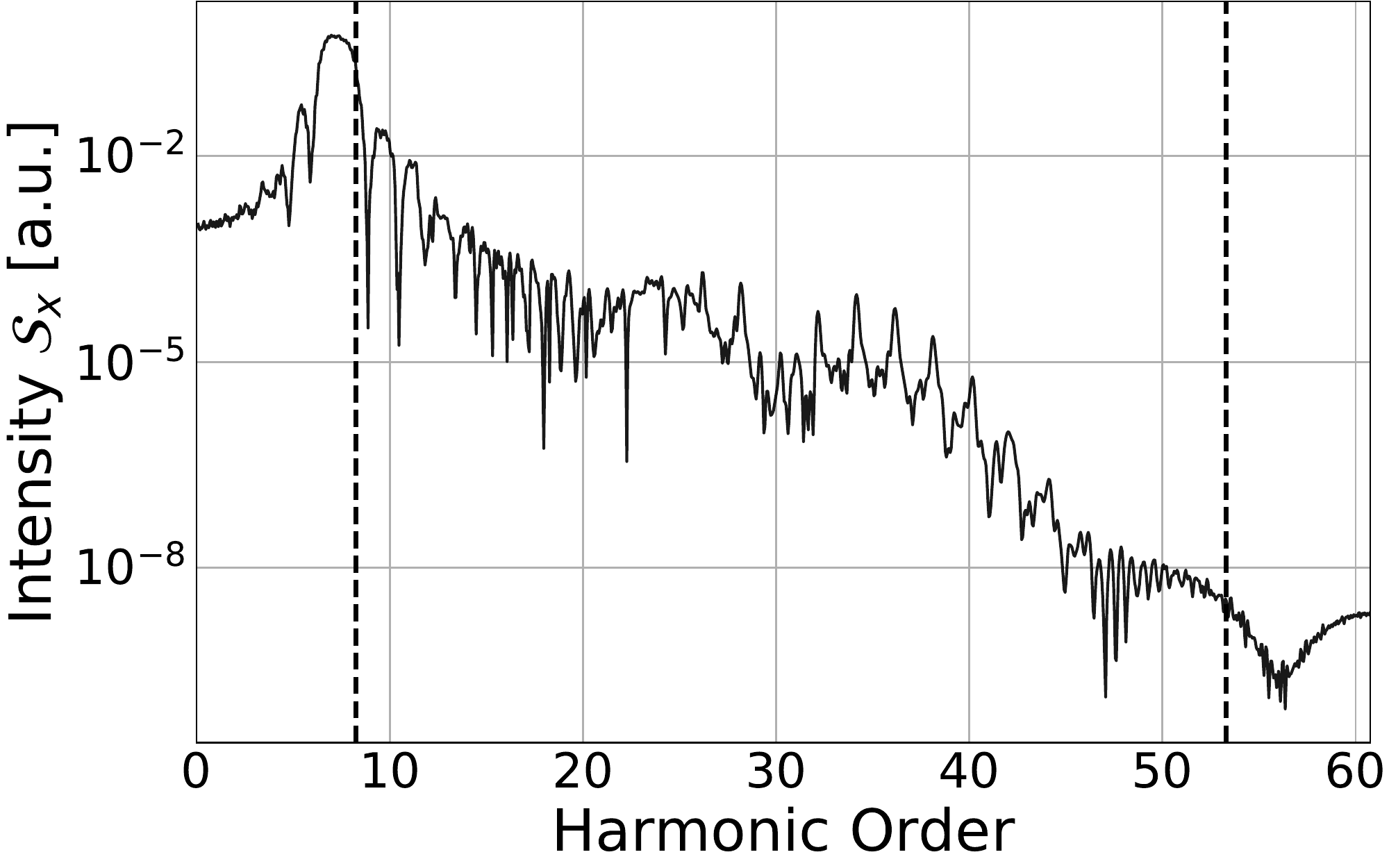}
	\end{center}
	\caption{Transverse $x$-component of spectrum of the HHG emission from an initially neutral atom with $90\,\%$ population in the excited hydrogen state $\psi_{311,0}(\vec{r})$. The atomic system is ionized by a laser with electric field strength $E_0=0.1026$, frequency $\omega_0=0.057$ and an interaction time of six optical cycles, $T=7T_0$. 
	Black dashed lines show the standard HHG plateau cutoffs $I_p/\omega_0$ and $(I_p+3.17U_p)/\omega_0$, respectively.}
	\label{fig:spectrum_experiment}
\end{figure}

To illustrate this microscopic mechanism we assume the atom to be already prepared in a superposition of the ground state and the excited state $\psi_{nlm,b}(\vec{r})=\psi_{311,0}(\vec{r})$, namely $\psi(\vec{r}, t=0) = \sqrt{0.1} \, \psi_{100,0}(\vec{r}) + \sqrt{0.9} \, \psi_{311,0}(\vec{r})$. Subsequently, the system is driven by an $800\,\mathrm{nm}$ infra-red laser pulse with a two-cycle flat-top and two-cycle ramps at the beginning and end of the pulse. The numerically obtained emission spectrum (for details see Appendix~\ref{app:num_sim}) from recombination into the ground state with the usual $\ionpot + 3.17\pondpot$ cutoff is shown in Fig.~\ref{fig:spectrum_experiment}. 

\section{Summary}
In summary, we have investigated HHG generated from twisted electronic wavepackets with orbital angular momentum quantized along the laser polarization axis. For  the dominant recombination into the ground state we have provided analytically the  spectrum  which has the same cutoff as results from an unstructured wavepacket. We inferred that for wavepackets with single vortex charges the circularly polarized pulse emitted propagates along the polarization axis, perpendicular to the laser propagation in contrast to the linearly polarized emission perpendicular to the polarization of HHG without angular momentum transfer. 

Hence, this may provide a source for circularly polarized attosecond pulses with an alternative quantum pathway compared to previous generation schemes~\cite{Milosevic2000,Medisauskas2015,Dorney2017,Xie2008,Ferre2014,Fleischer2014,Kfir2014, HernandezGarcia2016, Yuan2011}. In contrast, these methods employ elliptical driving fields and achieve either only quasi-circular harmonics or circular polarization for only selected harmonics. Instead, the circular polarization is uniform across the high harmonic spectrum in the approach presented here and therefore suitable for the generation of attosecond pulses with well defined circular polarization. It remains to be studied if it can compete in conversion efficiency against the previous works as well.

To this end we propose an experiment which converts a weak circularly polarized pulse with the help of intense CW linearly polarized light into circularly polarized attosecond pulses of high frequency using a gas of simple atoms with s-state valence orbitals as the HHG medium.

Such an experiment would at the same time be able to confirm the mechanism we have identified here to be responsible for HH generation with twisted electrons.
\begin{acknowledgments}
We thank Ulf Saalmann for helpful discussions. JMR acknowledges support from the Deutsche Forschungsgemeinschaft  through Priority Programme 1840 ``Quantum Dynamics in Tailored Intense Fields'' (QUTIF). 
\end{acknowledgments}

\appendix

\section{High harmonic yield from other recombination channels}
\label{app:hhg_yield}
\begin{figure}
	\begin{center}
	\includegraphics[width=0.5\textwidth]{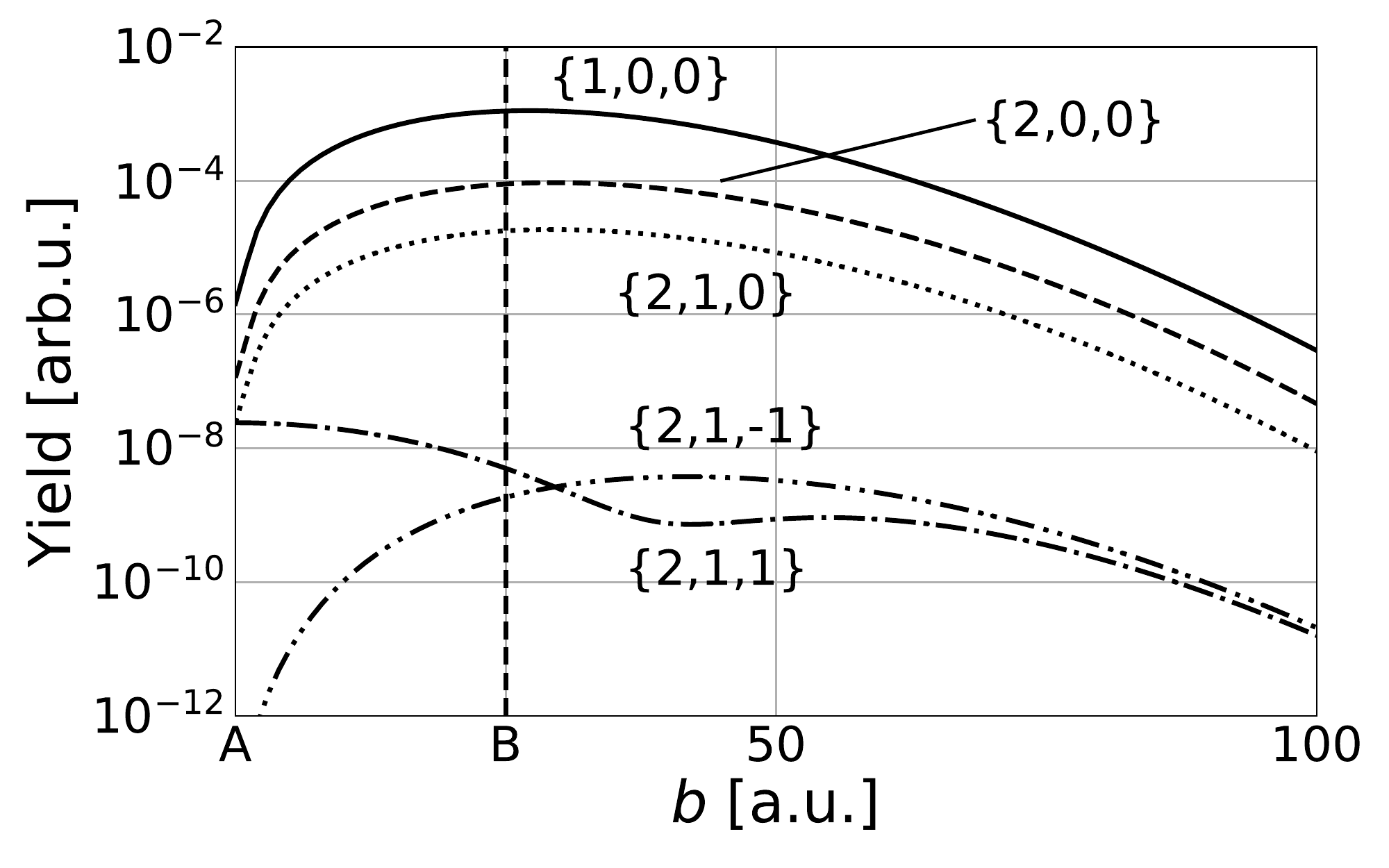}
	\end{center}
	\caption{HHG yield $Y_{b}^{(s=1)}$ for different recombination channels and varying impact parameter $b$. Different hydrogen eigenstates with quantum numbers $\{n,l,m\}$ are used. The initial transverse maxima $\rhomax^{(1)}(0)=\beta$ is indicated by the vertical, dashed line and the two scenarios A and B from Fig.~\ref{fig:twisted_topo} are marked on the horizontal axis.}
	\label{fig:yield_channels_s_1_groundstate}
\end{figure}

The HHG yields
\begin{equation}
		Y_{b}^{(s)} \equiv \int_{\wmin}^{\wmax} \dd{\omega} \mathcal{S}^{(s)}_{b}(w)
\end{equation}
for different recombination channels with $s=1$ are shown in Fig.~\ref{fig:yield_channels_s_1_groundstate}. The ground state recombination has the largest yield compared to other recombination channels by at least an order of magnitude at all impact parameters and validates the reduction to an effective two-level system consisting only of the twisted Volkov wavepacket and the unperturbed ground state.

\section{Finite length Fourier transformation}
\label{app:ft_filter}
\begin{figure}
	\begin{center}
	\includegraphics[width=0.3\textwidth]{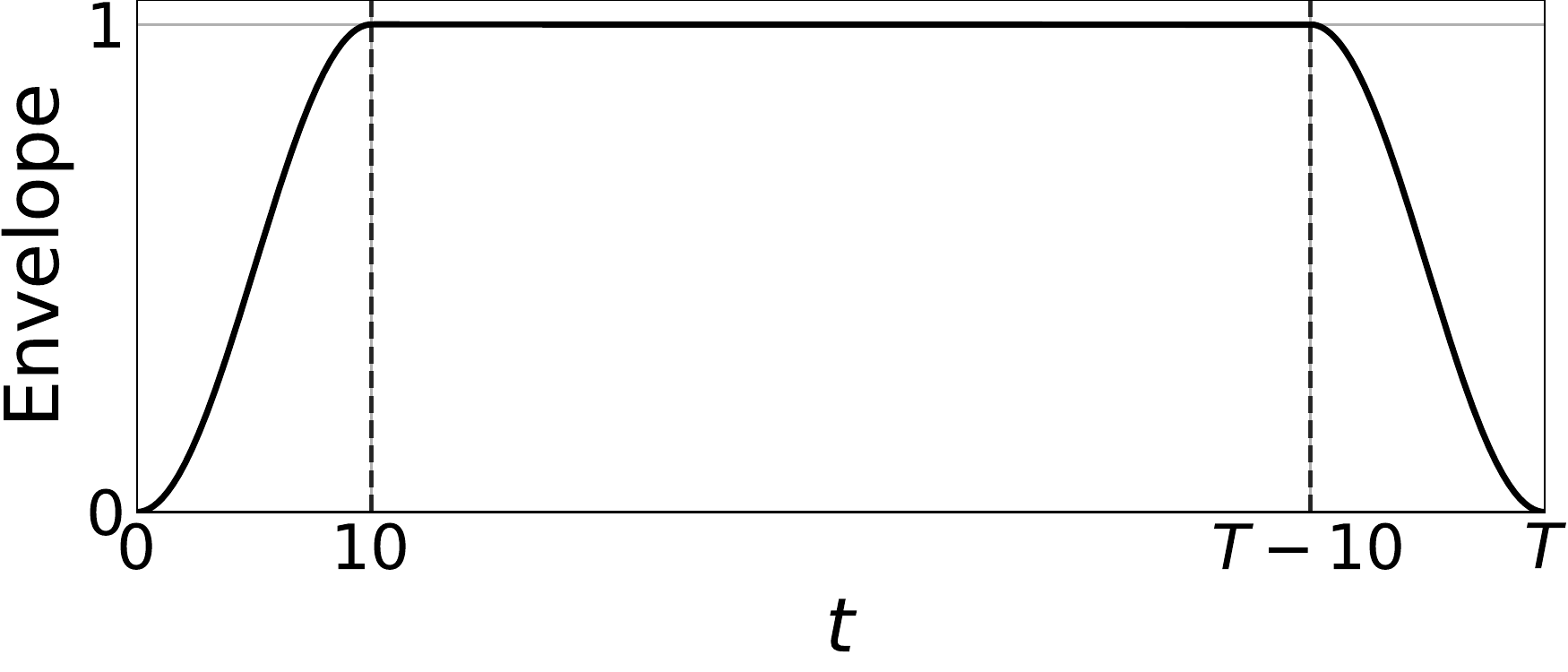}
	\end{center}
	\caption{Envelope function used for all Fourier transformations of signals with length $T$ in order to decrease finite length effects.}
	\label{fig:envelope}
\end{figure}

In order to decrease finite length effects from discrete Fourier transformations all signals were faded in and out with a quarter-period sine-square envelope over a time period of 10 atomic units. This is similar to the Tukey window~\cite{Harris1978} and the full envelope function is illustrated in Fig.~\ref{fig:envelope}.

\section{Numerical solution of Schr\"odinger equation}
\label{app:num_sim}
Due to numerical difficulties arising from the singularity of the Coulomb potential, we use a soft-core potential
\begin{equation}
	V_{c,b}^{(\alpha)}(\vec{r}) = - \frac{1}{( \abs{ \vec{r} - b\vu{x} }^4 + \alpha^4 )^{1/4} }
\end{equation}
with smoothening parameter $\alpha=0.2$. We approximate the ground state with a trial wavefunction
\begin{equation}
	\psi_{100,b=0}^{\text{sim}}(\vec{r},\aeff) = \frac{1}{ \sqrt{\pi \aeff^3} } \exp \left[ - \frac{r}{\aeff}  \right] \label{eq:trial_wf}
\end{equation}
and use the variational principle to estimate the ground state energy. We find a minimal energy $E^{\text{sim}}_{n=1} \approx -0.47$ for $\aeff \approx 1.086$. For our purpose it is sufficiently close to the real ground state and the trial wavefunction~(\ref{eq:trial_wf}) is used in the numerically obtained transition elements~(\ref{eq:transition_el}). We employ a split-operator method for the short-time evolution operator~\cite{Grossmann2013}
\begin{equation}
	\mathcal{U}(t+\Delta t, t) \approx e^{-\ii V(t+\Delta t/2)\Delta t/2} e^{-\ii T_k\Delta t} e^{-\ii V(t+\Delta t/2)\Delta t/2}
\end{equation}
with $V(t) \equiv V_{c,b}^{(\alpha)} + V_{A}(t) - \ii V_{\text{abs}}$ and $T_k \equiv -\vec{\nabla}_{\vec{r}}^2 / 2$ in Cartesian coordinates. Unwanted reflections at the grid boundaries are suppressed by a quadratic imaginary potential term $-\ii V_{\text{abs}}$ in the vicinity of the boundaries. The FFTW library~\cite{FFTW05} is used to transform the spatial wavefunction into momentum space for the evaluation of the kinetic energy term $e^{-\ii T\Delta t}$. 

\subsection{Scattering scenario}
\begin{figure}
\begin{center}
\begin{overpic}[width=0.5\textwidth,grid=False]{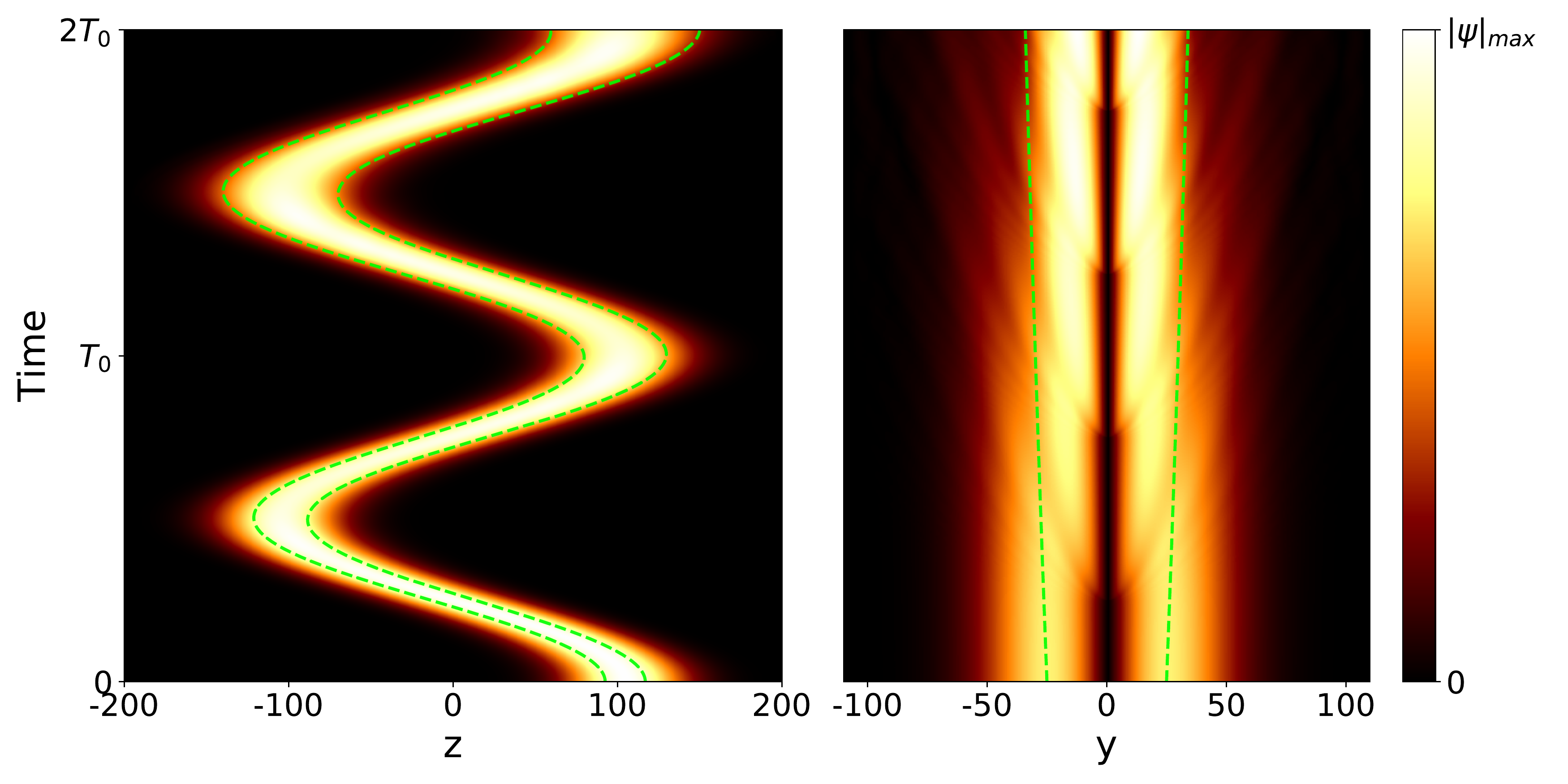}
	\put(9,44){\textcolor{white}{\textbf{(a)}}}
	\put(55,44){\textcolor{white}{\textbf{(b)}}}
\end{overpic}
\caption{(a) Longitudinal cross-section $\abs{\psi_{\text{sim}}(x=0,y=\beta,z)}$ of the numerically simulated wavefunction with analytical, longitudinal width (green dashed lines). (b) Transverse cross-section $\abs{\psi_{\text{sim}}(x=0,y,z=\volquiv)}$ of the numerically simulated wavefunction. The analytical, transverse maxima of Eq.~(\ref{eq:twisted_wp}) are indicated by the green dashed lines.}
\label{fig:focusing}
\end{center}
\end{figure}

For the scattering scenario, the initial wavefunction is taken to be the analytical form~(\ref{eq:twisted_wp}) of the twisted electron at time $t=0$ and is time-iterated under a laser pulse with electric field strength $E_0 = 0.05835$, frequency $\omega_0 = 0.02360$ and an interaction time of two optical cycles, $T=2T_0$. The specifications of the numerical grid are given in Table~\ref{tab:grid}. In Fig.~\ref{fig:focusing} we present the evolution of the electronic wavepacket and compare to the theoretical time-dependent width. The positions of the transverse maxima lie closer to the ion than the theoretical predictions and show the effect of the Coulomb potential on the twisted wavefunction.
\begin{table}
\caption{\label{tab:grid}Grid parameters for numerical simulation of scattering scenario.}
\begin{ruledtabular}
\begin{tabular}{ccccc}
Direction & Grid Points & Min & Max & Resolution $\Delta$ \\
\hline
x & $384$ & $-110$ & $110-\Delta x$ & $0.57$ \\
y & $384$ & $-110$ & $110-\Delta y$ & $0.57$ \\
z & $1536$ & $-200$ & $200-\Delta z$ & $0.26$ 
\end{tabular}
\end{ruledtabular}
\end{table}

\subsection{Experimental scenario with ionization process}
Assuming that a neutral hydrogen atom has been prepared in a superposition of the ground state with the excited state $\psi_{nlm,b}(\vec{r}) = \psi_{311,0}(\vec{r})$, we numerically evolve the initial state $\psi_0(\vec{r}) = \sqrt{0.1} \, \psi_{100,0} + \sqrt{0.9} \, \psi_{311,0}(\vec{r})$ under a laser pulse of the form $E(t) = E_0\, g(t)\, \sin(\omega_0 t + \varphi_0)$ with the envelope
\begin{equation}
	g(t) = \begin{cases}
		\sin^2\left(\frac{\pi}{2} \frac{t}{2T_0} \right), & 0 \leq t < 2T_0 \\
		1, & 2T_0 \leq t \leq 5T_0 \\
		\cos^2\left(\frac{\pi}{2} \frac{t - 4T_0}{2T_0} \right), & 5T_0 < t \leq 7T_0 \\
		0, & \mathrm{otherwise}
	\end{cases}
\end{equation}
and an initial phase $\varphi_0$. To emulate the widely available Ti:Sa laser, we use the optical frequency $\omega_0 = 0.057$ and the electric field strength $E_0 = 0.1026$. In contrast to the tunnel ionization of the ground state in HHG, the higher lying excited state leaves the Coulomb potential mainly via over-the-barrier ionization and is therefore sensitive to the initial phase of the driving field. We choose $\varphi_0 = 0.28\pi$ for only small drifts away from the core. The emission spectrum is obtained via the absolute square value of the Fourier transform of the dipole acceleration $\vec{a}(t) \equiv \mel{\psi(t)}{-\nabla V_{c,b=0}^{(\alpha)}}{\psi(t)}$. The grid specifications are given in Table~\ref{tab:grid_exp}.

\begin{table}
\caption{\label{tab:grid_exp}Grid parameters for numerical simulation including ionization.}
\begin{ruledtabular}
\begin{tabular}{ccccc}
Direction & Grid Points & Min & Max & Resolution $\Delta$ \\
\hline
x & $384$ & $-75$ & $750-\Delta x$ & $0.39$ \\
y & $384$ & $-75$ & $750-\Delta y$ & $0.39$ \\
z & $1920$ & $-250$ & $250-\Delta z$ & $0.21$ 
\end{tabular}
\end{ruledtabular}
\end{table}

\bibliography{hhg_twisted_electrons_bib}

\end{document}